\documentstyle[pra,preprint,aps]{revtex}

\tighten

\renewcommand{\baselinestretch}{1}
\begin{document}
\title{Quantum deleting and Signalling}
\author{Arun K. Pati$^{(1,2)}$ and Samuel L. Braunstein $^{(2)}$}
\address{$^{(1)}$ Institute of Physics, Bhubaneswar-751005, Orissa, India}
\address{$^{(2)}$ School of Informatics, Dean Street, University of Wales, Bangor LL 57 
1UT, U.K.}

\date{\today}
\maketitle
\def\ra{\rangle}
\def\la{\langle}
\def\ver{\arrowvert}
\begin{abstract}
It is known that if we can clone an arbitrary state we can send signal faster
than light. Here, we show that deletion of unknown quantum state against 
a copy can lead to superluminal signalling. But erasure of unknown quantum 
state does not imply faster than light signalling.
\end{abstract}



\vskip 1cm


\vskip 1cm



\par

Two deep  rooted concepts in  quantum theory are  linear superposition
principle  and   linear  evolution  equations   \cite{dirac}.   Linear
superposition principle  is the one  which makes any two  state quantum
system  a  unique  one, namely,  a  qubit  which  is not  realized  in
classical world.  The  real power of qubits is  being exploited in the
emerging  field  of  quantum  computation and  information  technology
\cite{mike}.   On   the  other  hand  linear   evolution  makes  certain
operations  impossible on arbitrary  superposition of  quantum states.
One of the simplest, yet  profound principle of quantum theory is that
we cannot clone \cite{wz,dd} an unknown quantum state exactly. This is
a  consequence  of  linearity  of  the evolution  of  quantum  states.
Subsequently, using  unitarity of quantum evolution it  was shown that
two   non-orthogonal  quantum  states   cannot  be   perfectly  copied
\cite{hy}. If we give up the  requirement of perfect copies then it is
possible  to  copy an  unknown  state  approximately by  deterministic
cloning  machines  \cite{bh,bbhb,bh1,nc,gm,bem}. Recent  understanding
suggests that non-orthogonal states  from a linear independent set can
be copied  exactly by a unitary and  measurement process \cite{dg,dg1}
and can  evolve into  a linear superposition  of multiple  copy states
\cite{akp}.

But what  could go wrong  if we can  clone an arbitrary state?  It was
already  known before  no-cloning theorem  that  if one  can clone  an
arbitrary   state    then   using   non-local    resources   such   as
Einstein-Podolsky-Rosen  (EPR) pair  one can  send signal  faster than
light \cite{nh}.   No-cloning theorem came to rescue  the violation of
causality  in the same  year \cite{wz,dd}.  Thus, linear  evolution of
quantum theory  and principle of  special theory of relativity  are in
peace. In  fact, one can  go a step  further and ask  if no-signalling
condition  can give  basic axiomatic  structures of  quantum mechanics
\cite{gsv}.   The fidelity  of universal  cloning (both  symmetric and
asymmetric) allowed  by quantum theory and  no-signalling condition of
special relativity are at just boarder line of crossing each other, in
the sense that if one could  have even a smallest departures from that
value, one  could send signals faster than  light \cite{ng,gkr}. Also,
it was shown  that even with a probabilistic  exact cloning machine we
cannot send superluminal signals \cite{akp1}.

Recently, it was proved that given two copies of an unknown quantum
state we cannot delete a copy against the other by any physical
operation (trace preserving completely positive
transformation)---called `no-deletion theorem' \cite{pb}.  This is,
yet, another fundamental consequence of linearity of quantum theory
and not restricted to class of operations such as unitaries.  The
deletion of quantum information should not be confused with the
erasure.  In the erasure of quantum information we can get rid of the
very last qubit (also called primitive erasure) by an irreversible
manner.  As in the classical case, erasure process always involves
spending certain amount of energy \cite{ls,rl} whereas deleting refers
to a un-copying type of operation.  The essential difference between
classical and quantum is, in classical world deletion and erasure both
are allowed but in quantum world deletion is not allowed whereas
erasure is.  Quantum no-deleting principle has also been generalized
to higher dimensional quantum systems and also for non-orthogonal
states using unitarity in a conditional manner  \cite{pb1} . Even
though one cannot delete two non-orthogonal states perfectly using
only unitary operation one can delete them in a probabilistic manner
\cite{feng}.  Surprisingly, a recent result states that to copy a
state from a non-orthogonal set the full information about  the clone
must already be provided in the ancilla state which is called a
`stronger no-cloning theorem' \cite{rj}. It is also suggested that the
stronger no-cloning and no-deleting theorems taken together provide a
property of `permanence' to quantum information.

In this letter we ask the question suppose one can delete an arbitrary
state using a  quantum deleting machine, then what  could go wrong? We
show that  if one can delete  unknown states then one  can send signal
faster  than light!  At  first glance  this  may be  surprising as  by
deleting  information  we are  only  reducing  the  redundancy at  our
disposal  and that  should  not affect  in  any way  the signal  being
sent. But on the other hand,  we know that linearity of quantum theory
has  survived to  its  highest precession  test  and a  little bit  of
non-linearity  would  allow  superluminal  signalling.  Then  one  can
perhaps  comprehend  that  since   no-deleting  is  a  consequence  of
linearity then any process that  violates linearity can clash with one
of  the  corner stone  of  special  theory  of relativity.  Therefore,
deletion of arbitrary state  could lead to signalling. Furthermore, we
show that erasure of quantum  information does not imply signaling, as
expected.

First, let us recall the quantum no-deletion principle.
Consider two copies of an unknown qubit $\arrowvert \psi \rangle$
each in a Hilbert space ${\cal H} = {\cal C}^{2}$.
Two copies live in a smaller dimensional subspace, which is the 
symmetric subspace of ${\cal H}{\otimes }{\cal H}$. It contains states that 
are symmetric under interchange of any pair of qubits. 
Quantum no-deleting principle states that it is impossible to design a 
machine that can delete one copy from a collection of two 
copies of unknown quantum state. That is there is no linear transformation 
${\cal L}:  {\cal H}_1\otimes {\cal H}_2 \otimes {\cal H}_3 \rightarrow 
 {\cal H}_1 \otimes {\cal H}_2 \otimes {\cal H}_3$
that will take
\begin{eqnarray}
\arrowvert \psi \rangle_1\arrowvert \psi \rangle_2 \arrowvert A \rangle_3 
\rightarrow \arrowvert \psi \rangle_1 \arrowvert\Sigma \rangle_2 
\arrowvert A_{\psi} \rangle_3,
\end{eqnarray}
where $\arrowvert\Sigma \rangle$ is the blank state which can be of our 
choice, $ \arrowvert A \rangle$ is the initial
and $\arrowvert A_{\psi} \rangle$ is the final state of the ancilla 
independent of $\arrowvert \psi \rangle$ (this is necessary to exclude 
swapping). It was shown that linearity does not allow us to delete an unknown 
state against a copy except swapping the unknown state onto the Hilbert
space of ancilla. However, the later operation is not a proper 
deletion as one can retrieve the original from the ancilla \cite{pb,pb1}.

To show that deletion of arbitrary state implies signalling consider the
following scenario.
Let Alice and Bob share two pairs of EPR singlets 
$\arrowvert\Psi^{-} \rangle_{12}$ and $\arrowvert\Psi^{-} \rangle_{34}$
and they are located in remote places. (Note that in proving cloning implies 
signalling one needs a single EPR pair to be shared between Alice and Bob.) 
Alice has particles $1$ and $3$ and Bob has $2$ and $4$. Since the singlet 
state is invariant under local unitary operation $U_i \otimes U_j, (i=1,3) 
(j =2,4)$ it is same in all basis (up to ${\rm U}(1)$ phase factors).
Let us write the combined state of the system in an arbitrary qubit basis
$\{ \arrowvert\psi \rangle = \cos \theta/2 \arrowvert 0 \rangle +
\sin \theta/2 \arrowvert 1 \rangle, \arrowvert{\bar \psi} \rangle =
\sin \theta/2 \arrowvert 0 \rangle - \cos \theta/2 \arrowvert 1 \rangle \}$
as                              
\begin{eqnarray}
\arrowvert\Psi^{-} \rangle_{12} \arrowvert\Psi^{-} \rangle_{34} &=&
\frac{1}{2}\bigg(\arrowvert\psi \rangle_1 \arrowvert\psi \rangle_3
\arrowvert{\bar \psi} \rangle_2 \arrowvert{\bar \psi} \rangle_4 +
\arrowvert{\bar \psi} \rangle_1 \arrowvert{\bar \psi} \rangle_3 
\arrowvert\psi \rangle_2 \arrowvert\psi \rangle_4  \nonumber\\
&-& \arrowvert{\bar \psi} \rangle_1 \arrowvert\psi \rangle_3 
\arrowvert\psi \rangle_2 \arrowvert{\bar \psi} \rangle_4
- \arrowvert\psi \rangle_1 \arrowvert{\bar \psi} \rangle_3 
\arrowvert{\bar \psi} \rangle_2 \arrowvert\psi \rangle_4 \bigg).
\end{eqnarray}

First we notice that the state of the Bob's particles $2$ and $4$ are in a 
completely random mixture. Now Alice can measure her particles $1$ and $3$ 
onto the qubit basis
$\{ \arrowvert \psi \rangle, \arrowvert{\bar \psi} \rangle \}$. If the outcome
is $\arrowvert\psi \rangle_1 \arrowvert\psi \rangle_3$, then after 
communicating the result to Bob, Bob's
particles $2$ and $4$ are in the state $\arrowvert{\bar \psi} \rangle_2
\arrowvert{\bar \psi} \rangle_4 $. If Alice's outcome is 
$\arrowvert{\bar \psi} \rangle_1 \arrowvert{\bar \psi} \rangle_3$, then
after receiving classical communication Bob's particles are in the state 
$\arrowvert\psi \rangle_2 \arrowvert\psi
\rangle_4$. Similarly, one can find the resulting states with other choices of
measurements. However, whatever measurements Alice does, if she does not convey
the measurement results to Bob, then Bob's particles are in a completely
random mixture, i.e. $\rho_{24} = \frac{I_2}{2} \otimes \frac{I_2}{2}$.
That is to say that local operations on Alice's Hilbert space 
${\cal H}_1\otimes {\cal H}_3$ has no effect on the Bob's description 
of the state in the Hilbert space ${\cal H}_2\otimes {\cal H}_4$.
As is well known, the result of any measurement (von Neumann or POVM) that Bob will perform
on his particles will depend only on the reduced density matrix of the particle
$2$ and $4$.

But suppose Bob has a quantum deleting machine which can delete
an arbitrary state. The action of quantum deleting machine 
on the two copies and the ancilla state belonging to the Hilbert space 
${\cal H}_2 \otimes {\cal H}_4 \otimes {\cal H}_5$  
can be described by 
\begin{eqnarray}
&& \arrowvert\psi \rangle_2 \arrowvert \psi \rangle_4 \arrowvert A \rangle_5
\rightarrow \arrowvert \psi \rangle_2 \arrowvert\Sigma \rangle_4  
\arrowvert A_{\psi} \rangle_5 \nonumber \\
&& \arrowvert{\bar \psi} \rangle_2 \arrowvert{\bar \psi} \rangle_4 
\arrowvert A \rangle_5
\rightarrow \arrowvert{\bar \psi} \rangle_2 \arrowvert\Sigma 
\rangle_4  \arrowvert 
A_{{\bar \psi}} \rangle_5 \nonumber \\
&& \arrowvert\psi \rangle_2 \arrowvert{\bar \psi} \rangle_4 \arrowvert A 
\rangle_5  \rightarrow \arrowvert \phi' \rangle_{245}  \nonumber \\
&& \arrowvert{\bar \psi} \rangle_2 \arrowvert\psi \rangle_4 \arrowvert A 
\rangle_5 \rightarrow \arrowvert \phi'' \rangle_{245}.
\end{eqnarray}
The last two transformations correspond to the situation when the states are
non-identical and in these cases the output state can be some arbitrary
entangled states, in general.
After passing through the quantum deleting machine the combined state of 
Alice and Bob transforms as
\begin{eqnarray}
\arrowvert\Psi^{-} \rangle_{12} \arrowvert\Psi^{-} \rangle_{34}
\arrowvert A \rangle_5 \rightarrow
\frac{1}{2} \bigg(\arrowvert\psi \rangle_1 \arrowvert\psi \rangle_3
\arrowvert{\bar \psi} \rangle_2 \arrowvert \Sigma \rangle_4 
\arrowvert A_{{\bar \psi}} \rangle_5 +
\arrowvert{\bar \psi} \rangle_1 \arrowvert{\bar \psi} \rangle_3 
\arrowvert\psi \rangle_2 \arrowvert \Sigma \rangle_4 
\arrowvert A_{\psi} \rangle_5  \nonumber \\
- \arrowvert{\bar \psi} \rangle_1 \arrowvert\psi \rangle_3 
\arrowvert\phi' \rangle_{245}
- \arrowvert\psi \rangle_1 \arrowvert{\bar \psi} \rangle_3 
\arrowvert \phi''\rangle_{245} \bigg) = |\Psi^{(\rm out)}\rangle_{12345}.
\end{eqnarray}

Suppose Alice and Bob have pre-agreed that the measurements onto
basis states $\{|0\rangle,|1\rangle\}$ means $`0'$ and onto 
any other (say) $\{|\psi \rangle,|{\bar \psi} \rangle\}$ means $`1'$.
Now, Alice performs measurements onto either of these two choices of
basis states but does not communicate the measurement outcome.
Since Bob is ignorant of Alice's measurement, he traces out the particles 
at Alice's lab and the ancilla at his lab too. The reduced density
matrix for particles $2$ and $4$ at Bob's place is given by

\begin{eqnarray}
&& \rho_{24} = {\rm tr}_{135} \bigg[\rho_{12345}^{(\rm out)} \bigg]  = 
\frac{1}{4}\bigg[ I_2 \otimes
\arrowvert \Sigma \rangle_4{_4}\langle \Sigma \arrowvert +
\rho_{24}' + \rho_{24}'' \bigg],
\end{eqnarray}
where $\rho_{12345}^{(\rm out)} = |\Psi^{(\rm out)} \rangle_{12345} 
\langle \Psi^{(\rm out)}|$, $\rho_{24}' =
{\rm tr}_5 (\arrowvert \phi' \rangle_{245}\langle \phi' \arrowvert)$
and $\rho_{24}'' =
{\rm tr}_5 (\arrowvert \phi'' \rangle_{245} \langle \phi'' 
\arrowvert)$.
Since $\arrowvert \phi' \rangle_{245}$ and $\arrowvert \phi'' \rangle_{245}$ 
are in general pure entangled states of the non-identical inputs and the 
ancilla, it will depend on the input parameters. After tracing out the 
ancilla, we will have, in general, mixed entangled states given by
\begin{eqnarray}
&& \rho'_{24}(\theta) =
\frac{1}{4}\bigg(I_2\otimes I_4 + {\bf m}'(\theta).\sigma_2 \otimes I_4 +
I_2 \otimes {\bf n}'(\theta).\sigma_4 + \sum_{ij} C_{ij}(\theta)'
{\sigma_i}_2 \otimes {\sigma_j}_4 \bigg) \nonumber\\
&& \rho''_{24}(\theta) = 
\frac{1}{4} \bigg(I_2\otimes I_4 + {\bf m}''(\theta).\sigma_2 \otimes I_4 +
I_2 \otimes {\bf n}''(\theta).\sigma_4 + \sum_{ij} C_{ij}''(\theta)
{\sigma_i}_2 \otimes {\sigma_j}_4 \bigg).
\end{eqnarray}
Thus, it is clear that the reduced density matrix of particles $2$ and
$4$ at Bob's place are no more completely random and it depends on the
choice of basis.  This shows that if Alice measures her particles in
$\{ \arrowvert 0 \rangle,  \arrowvert1 \rangle\}$ basis then the
density matrix of Bob's particles will be in $\rho_{24}(0)$.  If Alice
measures her particles in $\{ \arrowvert \psi \rangle,  \arrowvert
{\bar \psi} \rangle\}$ basis then Bob's particles will be described
by  a different density matrix $\rho_{24}(\theta)$. Since these two
statistical mixtures are non-identical Bob can  distinguish
them. Therefore, by {\em deleting an arbitrary state } he can
distinguish two statistical mixtures  and that {\em will allow
communication of one classical bit superluminally }. It is known that
if one allows non-linear operation one can distinguish two statistical
mixtures \cite{gisin}. This suggests that possibly the action of
deletion was a non-linear operation beyond the realm of quantum theory.

Furthermore, we can show that erasure of unknown state does not 
imply superluminal
signalling. In a stronger form, erasure of information can be accomplished by swapping the last
qubit with a standard state and then dumping it into the environment. Suppose
Bob is performing erasure operation on his particles at his disposal. In this
case, Bob can simply choose the initial state of the ancilla $\arrowvert A 
\rangle$ to be the blank state $\arrowvert \Sigma \rangle$. Then, he performs
swapping of the last two qubits in ${\cal H}_4\otimes {\cal H}_5$ and tracing 
over Hilbert space ${\cal H}_5$. Now instead of transformation (3) we have 
\begin{eqnarray}
&& \arrowvert\psi \rangle_2 \arrowvert \psi \rangle_4 \arrowvert 
\Sigma \rangle_5 \rightarrow \arrowvert \psi \rangle_2 \arrowvert 
\Sigma \rangle_4 \arrowvert \psi \rangle_5 \nonumber \\
&& \arrowvert{\bar \psi} \rangle_2 \arrowvert{\bar \psi} \rangle_4 
\arrowvert \Sigma \rangle_5
\rightarrow \arrowvert{\bar \psi} \rangle_2 \arrowvert \Sigma 
\rangle_4  \arrowvert {\bar \psi} \rangle_5 \nonumber \\
&& \arrowvert\psi \rangle_2 \arrowvert{\bar \psi} \rangle_4 \arrowvert \Sigma 
\rangle_5
\rightarrow \arrowvert \psi \rangle_2 \arrowvert \Sigma 
\rangle_4  \arrowvert {\bar \psi} \rangle_5 \nonumber \\
&& \arrowvert{\bar \psi} \rangle_2 \arrowvert\psi \rangle_4 
\arrowvert \Sigma \rangle_5 
\rightarrow \arrowvert{\bar \psi} \rangle_2 \arrowvert \Sigma 
\rangle_4  \arrowvert \psi \rangle_5 .
\end{eqnarray}
Using the argument as before, without any communication from Alice to Bob,
the two particle density matrix at Bob's place (after swapping and tracing over
the ancilla) is given by
\begin{eqnarray}
&& \rho_{24} =  \frac{I_2}{2} \otimes
\arrowvert \Sigma \rangle_4{_4}\langle \Sigma \arrowvert .
\end{eqnarray}
This density matrix does not carry any information about Alice's
choice of basis as it is independent of the parameter $\theta$.
Therefore, by erasing the information Bob will not be able to know 
onto which basis Alice has performed the measurement. Thus, the erasure
of unknown state does not lead to superluminal signalling.

The no-deletion theorem is a consequence of linearity of
quantum theory.
We have  shown that
violation of  no-deletion theorem indeed  can lead to  superluminal
signalling  using  non-local  entangled  states. However,  erasure  of
information  does not  imply  any signalling.  These two  observations
further illustrate the fact that the quantum deletion is fundamentally
a different operation than  the erasure.   

We end with a remark that classical information  is
physical but has no  permanence. {\em Quantum information is physical and
has permanence } (in view of recent stronger no-cloning and no-deleting theorems  
in quantum information \cite{rj}). Here, permanence refers to the fact that 
to duplicate quantum information, it must exist somewhere in the universe and to 
eliminate, it must be moved to somewhere else in the universe where it still exists.
It would be interesting to  see if the violation of permanence property of
quantum information can lead to superluminal signalling.
That it should be true is seen here partly (via deleting implies signalling). 
In future, it remains to be seen if negating stronger no-cloning theorem leads to signalling. 

\vskip  1cm  
\noindent
{\bf Acknowledgments:}
AKP thanks  useful  discussions  at  various stages  with
N. Cerf, A.  Chefles and  N. Gisin  on these  issues  during academic
visits in 2000. Financial support from ESF for some of these visits is
gratefully acknowledged.

\vskip 1cm

\renewcommand{\baselinestretch}{1}


\end{document}